\documentclass[10pt]{article}
\usepackage{amssymb}
\newtheorem{definition}{Definition}
\newtheorem{proposition}[definition]{Proposition}
\newtheorem{lemma}[definition]{Lemma}
\newtheorem{theorem}[definition]{Theorem}

\def\proof{{\sc Proof}}

\def\QED{$\Box$}

\def\id{\mathit{id}}
\def\iif{{\mathit{if}}}
\def\tthen{{\mathit{then}}}
\def\eelse{{\mathit{else}}}
\def\Fst{{\mathit{fst}}}
\def\Snd{{\mathit{snd}}}
\def\PCF{{\mathrm{PCF}}}
\def\lev{{\mathrm{lv}}}

\newcommand\pure[1]{\overline{#1}}
\newcommand\num[1]{\widehat{#1}}
\newcommand\ang[1]{\langle #1 \rangle}
\newcommand\dang[1]{\ll #1 \gg}
\def\ccase{{\mathtt{case}}}
\def\oof{{\mathtt{of}}}
\newcommand\caseof[2]{\ccase\;#1\;\oof\;(#2)}

\def\LLL{{\mathcal{L}}}
\def\N{\mathbb{N}}
\def\bool{{\mathtt{B}}}
\def\unit{{\mathtt{U}}} 
\def\nat{{\mathtt{N}}}
\def\obs{{\mbox{\scriptsize \rm obs}}}
\def\SP{{\mathsf{SP}}}

\def\iso{\cong}
\def\arrow{\rightarrow}
\def\pseudo{\trianglelefteq}
\def\isplit{\lessdot}
\def\restrict{\upharpoonleft}
\def\parrow{\rightharpoonup}
\def\reducesto{\rightsquigarrow}
\def\darrow{\Rightarrow}

\begin{document}

\title{The encodability hierarchy for PCF types}
\author{John Longley}
\maketitle

\begin{abstract}
Working with the simple types over a base type of natural numbers (including product types), 
we consider the question of when a type $\sigma$ is encodable as a definable retract of $\tau$: 
that is, when there are $\lambda$-terms $e:\sigma\arrow\tau$ and $d:\tau\arrow\sigma$ with 
$d \circ e = \id$.
In general, the answer to this question may vary according to both the choice of $\lambda$-calculus
and the notion of equality considered; 
however, we shall show that the encodability relation $\preceq$ between types
actually remains stable across a large class of languages and equality relations, ranging from a very
basic language with infinitely many distinguishable constants $\num{0},\num{1},\ldots$
(but no arithmetic) considered modulo computational equality,
up to the whole of Plotkin's PCF considered modulo observational equivalence.
We show that $\sigma \preceq \tau \preceq \sigma$ iff $\sigma \iso \tau$ via trivial isomorphisms, 
and that for any $\sigma,\tau$ we have either $\sigma \preceq \tau$ or $\tau \preceq \sigma$.
Furthermore, we show that the induced linear order on isomorphism classes of types is actually a well-ordering
of type $\epsilon_0$, and indeed that there is a close syntactic correspondence between simple types and
Cantor normal forms for ordinals below $\epsilon_0$.
This means that the relation $\preceq$ is readily decidable, and that
terms witnessing a retraction $\sigma \lhd \tau$ are readily constructible when $\sigma \preceq \tau$ holds.
\end{abstract}

\section{Introduction}

Consider the simple types generated by 
\[ \sigma,\tau ~::=~ \nat ~\mid~ \sigma\arrow\tau ~\mid~ \sigma\times\tau \]
where we take $\arrow$ to be right-associative and $\times$ to be left-associative,
and we think of $\nat$ as the type of natural numbers.

Loosely speaking, we shall be interested in the question:
when can a type $\sigma$ be \emph{encoded} in a type $\tau$? 
In other words, for which pairs of types 
$\sigma,\tau$ can one provide an `encoding' operation $e:\sigma\arrow\tau$ and a `decoding' operation
$d:\tau\arrow\sigma$ such that $d \circ e = \id_\sigma$? 
If such operations exist, one may say in mathematical terminology that $\sigma$ is a \emph{retract}
of $\tau$, with $e,d$ constituting a \emph{retraction} $\sigma\lhd\tau$.

For example, under mild assumptions, we can encode $\nat\times\nat$ in $\nat\arrow\nat$:
take an encoding $e$ that maps a pair $\ang{m,n}$ to the function 
$\lambda j.\,\iif\;j=0\;\tthen\;m\;\eelse\;n$,
and a decoding $d$ that maps a function $f$ to the pair $\ang{f(0),f(1)}$.
However, one would not expect to be able to encode $\nat\arrow\nat$ in $\nat\times\nat$
in a similar fashion.

To make our question precise, we have to clarify two things:

\begin{itemize}
\item What do we mean by an `operation' of type $\sigma\arrow\tau$ or $\tau\arrow\sigma$?
One possibility is to take this to mean a closed term of the appropriate type in some simply-typed
$\lambda$-calculus (with product types), taking $\id_\sigma$ to be the term $\lambda x^\sigma.x$.
A minimal choice would be the pure simply-typed $\lambda$-calculus itself; we shall denote this by
$\LLL_0$. A slightly less minimal choice would be the simply-typed $\lambda$-calculus with constants
\[ \num{n} ~:~\nat \;, ~~~~~~~~  
    \iif_{n} ~:~\nat\arrow\nat\arrow\nat\arrow\nat  \mbox{~~~for each $n \in \N$,} \]
subject to the conversion rules
\[ \iif_n\;\num{n}\;P\;Q ~=~ P \;, ~~~~~~~~
   \iif_n\;\num{m}\;P\;Q ~=~ Q \mbox{~~for each $m \neq n$.} \]
We shall denote this language by $\LLL_1$.
A much more generous choice would be the whole of Plotkin's PCF 
(suitably formulated with a single base type $\nat$ and with product types). 
Even richer languages might also be considered.%
\footnote{Alternatively, one could construe an `operation' to mean an element of the appropriate type within 
some \emph{model} of simply-typed $\lambda$-calculus. We shall not emphasize this `semantic' point of
view in this note, although it is clearly related to the syntactic one, as there are often close relationships
between particular $\lambda$-calculi and particular models.}

\item What does the `$=$' mean in the equation $d \circ e = \id_\sigma$?
For instance, a very strict kind of equality would be \emph{computational equality}, the congruence on
$\lambda$-terms generated by the conversion rules of the language in question.
We shall write $=_0$ for computational equality on $\LLL_0$ (this is generated by the $\beta$-rule plus
the equations $\Fst\,\ang{M,N} = M$, $\Snd\,\ang{M,N}=N$),
and $=_1$ for computational equality on $\LLL_1$ (where the conversion rules for $\iif_n$ are added). 
A much looser kind of equality would be some kind of \emph{observational equivalence} of terms,
such as the familiar notion of observational equivalence in $\PCF$; we shall denote this by $=_\obs$.
\end{itemize}

We shall refer to a choice of a simply-typed $\lambda$-calculus $\LLL$ 
together with an equality relation $=$ on its terms as a \emph{language theory}.
Once a language theory $(\LLL,=)$ has been fixed on, it becomes a precise question
for which $\sigma,\tau$ we have $\sigma\lhd\tau$;
we may thus write $\sigma \lhd_{(\LLL,=)} \tau$ to mean that the language theory $(\LLL,=)$
makes $\sigma$ a retract of $\tau$.
However, one can imagine that the relation $\lhd_{(\LLL,=)}$ might vary according to the 
language theory chosen: in principle, the richer the language, and the more generous
the equality, the easier it might become to construct a retraction $\sigma\lhd\tau$.
More precisely, suppose we set $(\LLL,=) \sqsubseteq (\LLL',=')$ iff $\LLL \subseteq \LLL'$ and
$= \;\subseteq\; =' \restrict_\LLL$;
in this situation, it is clear that $\sigma \lhd_{(\LLL,=)} \tau$
implies $\sigma \lhd_{(\LLL',=')} \tau$, though the converse need not hold in general.
It is thus not initially obvious whether any `stable' or `robust' answer 
to the question raised at the outset should be expected.

There has been a body of previous work on characterizing $\lhd_{(\LLL,=)}$ in the case 
$\LLL = \LLL_0$, both with respect to $\beta$-equality \cite{Bruce-Longo} and
more non-trivially with respect to $\beta\eta$-equality 
\cite{Liguoro-Piperno-Statman,Schubert,Padovani,Stirling}.
Such questions have also been considered in the presence of multiple base types
\cite{Regnier-Urzyczyn,Stirling}.
Closely related to this is a body of work on characterizing the \emph{isomorphism} relation between types
in pure typed $\lambda$-calculi, often with much richer type systems than the one considered here
(see \cite{Di-Cosmo-survey} for an informative survey).
It is also the case that for languages with the power of $\LLL_1$ or above,
particular examples of definable retractions arise routinely used in higher-order computability theory:
for example, one frequently exploits the fact that every simple type is encodable in a pure type
(see \cite[Chapter~4]{HOC}).
However, as far we are aware, there has hitherto been no systematic attempt to map out the 
encodability relation for all types in languages of this kind.

Our purpose of this paper is to study the question of encodability in the setting of such languages.
We will show, in fact, that the relation 
$\lhd_{(\LLL,=)}$ remains stable across a significant class of language theories $(\LLL,=)$;
furthermore, this relation is easy to characterize syntactically and enjoys some very pleasing properties.
More specifically, we shall establish the following:
\begin{enumerate}
\item The relation $\lhd_{(\LLL,=)}$ is the same for all language theories $(\LLL,=)$ with 
\[ (\LLL_1, =_1) ~\sqsubseteq~ (\LLL,=) ~\sqsubseteq~ (\PCF,=_\obs) \;. \]
We henceforth write this relation on types as $\preceq$.
\item We have $\sigma \preceq \tau \preceq \sigma$ if and only if $\sigma,\tau$ are 
\emph{trivially isomorphic}, i.e.\ iff there is an isomorphism between them generated by 
canonical isomorphisms of type
\begin{eqnarray*}
(\rho\times\rho')\arrow\rho'' & \iso & \rho\arrow\rho'\arrow\rho'' \\
\rho\arrow(\rho'\times\rho'') & \iso & (\rho\arrow\rho')\times(\rho\arrow\rho'') \\
\rho\times(\rho'\times\rho'') & \iso & (\rho\times\rho')\times\rho'' \\
\rho\times\rho' & \iso & \rho'\times\rho
\end{eqnarray*}
(This is a well-known axiomatization for isomorphisms of types built from $\arrow,\times$ in pure
$\lambda$-calculi: see e.g.\ \cite{Di-Cosmo-survey}.) 
It follows, for instance, that if $\sigma,\tau$ can be isomorphic relative to $(\PCF,\obs)$
(i.e. there are $\PCF$ terms $e:\sigma\arrow\tau$, $d:\tau\arrow\sigma$ with 
$d \circ e =_\obs \id_\sigma$, $e \circ d =_\obs \id_\tau$) only if they are trivially isomorphic.
\item For any $\sigma,\tau$, we have either $\sigma \preceq \tau$ or $\tau \preceq \sigma$.
Thus, $\preceq$ induces a total ordering (which we also write as $\preceq$) on types modulo
trivial isomorphism.
\item This total ordering on $\sim$-classes of types is in fact a well-ordering of order type $\epsilon_0$.
What is more, there is a close syntactic correspondence between simple types and
\emph{Cantor normal forms} for ordinals below $\epsilon_0$.
As we shall see, this correspondence leads to a simple syntactic characterization of $\preceq$
showing that this relation is readily decidable.
\end{enumerate}

We note at the outset that this picture may break if languages more powerful than $\PCF$ are admitted,
or if equalities more generous than $\PCF$ observational equivalence are considered.
On the one hand, if we move to a language such as PCF+parallel-or+exists or PCF+catch
for which a \emph{universal} type exists, then many other non-trivial encodings between types will be possible
\cite{HOC}.
(In an extension of PCF with higher-order references, even non-trivial \emph{isomorphisms} between types
can appear \cite{Clairambault-HO-refs}.)
On the other hand, if our language is $\PCF$ (or even System T) and we work up to equivalence
with respect to observing contexts drawn only from System T, we will find that every type actually becomes 
definably isomorphic to the pure type of the same level. 
This is an easy consequence of Theorem~4.2.9 of \cite{HOC},
which establishes this fact for extensional total type structures over $\N$ under mild hypotheses.

In Section~\ref{sec-positive} we establish the `positive' content of the above results: the existence of
an ordinal ranking on types leading to the definition of a total preorder $\preceq$ with associated equivalence
$\sim$; the existence of a trivial isomorphism whenever $\sigma \sim \tau$;
and the existence of a $(\LLL_1,=_1)$-retraction $\sigma\lhd\tau$ whenever $\sigma \preceq \tau$.
In Sections~\ref{sec-pure} and \ref{sec-negative} we proceed to the `negative' part, 
namely the fact that if $\sigma \not\preceq \tau$ then
no retraction $\sigma \lhd \tau$ exists, even with respect to $(\PCF,=_0)$.
We establish this using the technology of \emph{nested sequential procedures} for PCF.
Since the argument in full generality is quite complex, we first treat the case when $\tau$ 
is a pure type $\pure{k}$, then use this to motivate some of the ideas required for the general case.



I am grateful to Dag Normann, both for raising the question of characterizing the encodability relation
for all simple types in the setting of $\PCF$, and also for the key insight that deeply nested constituents
of types contribute more to their complexity than shallow ones: e.g.\
$(\nat^2 \arrow \nat) \arrow \nat$ is a more complex type than $(\nat \arrow \nat)^2 \arrow \nat$,
which is more complex than $((\nat \arrow \nat) \arrow \nat)^2$.
This was the idea that led to the ordinal ranking of types as exhibited in Section~\ref{sec-positive}.




\section{An ordinal ranking for types}  \label{sec-positive}

Let us begin by defining the relation $\sim$ on types to be the congruence
generated by the `trivial' equivalences mentioned above:
\begin{eqnarray*}
(\rho\times\sigma)\arrow\tau & \sim & \rho\arrow\sigma\arrow\tau \\
\rho\arrow(\sigma\times\tau) & \sim & (\rho\arrow\sigma)\times(\rho\arrow\tau) \\
\rho\times(\sigma\times\tau) & \sim & (\rho\times\sigma)\times\tau \\
\sigma\times\tau & \sim & \tau\times\sigma
\end{eqnarray*}
Clearly, each of these generating equivalences corresponds to an isomorphism of types
expressible in $(\LLL_0,=_0)$; it follows easily that 
if $\sigma \sim \tau$ then $\sigma \iso_(\LLL_0,=_0) \tau$.
If $\sigma \sim \tau$, we shall say that $\sigma,\tau$ are \emph{trivially isomorphic}.
We note in particular that $\rho \arrow \sigma \arrow \tau \sim \sigma \arrow \rho \arrow \tau$
for any $\rho,\sigma,\tau$, and that 
$\rho_0 \times \cdots \times \rho_{n-1} \sim \rho_{p(0)} \times\cdots\times \rho_{p(n-1)}$
for any permutation $p$ of $0,\ldots,n-1$.

To define $\preceq$ and the ordinal ranking of types, we shall work with the subclass of 
types $\rho,\theta$ generated by the grammar
\begin{eqnarray*}
\rho & ::= & \nat ~\mid~ \theta \arrow \nat \\
\theta  & ::= & \rho ~\mid~ \theta \times \rho
\end{eqnarray*}
We shall refer to these here as \emph{uncurried} types (ad hoc terminology).
It is easy to see that every $\sigma$ is trivially isomorphic to some uncurried type.

The following inductive clauses assign ordinal ranks $R(\rho), R(\theta) < \epsilon_0$ 
to certain well-behaved uncurried types:
\begin{itemize}
\item $R(\nat) = 1$.
\item If $R(\rho_1) \geq R(\rho_2) \geq \cdots \geq R(\rho_n)$ then
$R(\rho_1 \times \rho_2 \times\cdots\times \rho_n) = R(\rho_1) + R(\rho_2) + \cdots + R(\rho_n)$.
\item If $R(\theta)=\alpha$ then $R(\theta\arrow\nat) = \omega^\alpha$.
\end{itemize}
We may refer to the types $\rho,\theta$ to which a value for $R$ is assigned by this inductive definition as
\emph{canonical types}. (We might also add the empty product type $1$ and declare that $R(1)=0$,
but this would introduce complications later on which we prefer to avoid.)

Note that a canonical type $\sigma$ is in effect a representation of the \emph{Cantor normal form} 
for the ordinal $R(\sigma)$.
For our purposes, Cantor normal forms will be formal expressions generated inductively by the following clauses (we generate them simultaneously with a valuation $\nu$ mapping them to actual ordinals).
Note that we here modify the usual definition so as to exclude 0.
\begin{itemize}
\item $1$ is a Cantor normal form, where $\nu(1)=1$.
\item If $c$ is a Cantor normal form then so is $\omega^c$, where $\nu(\omega^c) = \omega^{\nu(c)}$.
\item If $c_0,\ldots,c_{n-1}$ are Cantor normal forms with 
$\nu(c_0) \geq \nu(c_1) \geq \cdots \geq \nu(c_{r-1})$, then $c_0 + \cdots + c_{r-1}$ is a Cantor normal
form with $\nu(c_0 + \cdots + c_{r-1}) = \nu(c_0) + \cdots + \nu(c_{r-1})$.
\end{itemize}
In practice, we shall sometimes blur the distinction between Cantor normal forms and the ordinals they denote.

The correspondence between Cantor normal forms and canonical types is now immediate.
The well-known fact that every ordinal below $\epsilon_0$ has a unique Cantor normal form now gives us:

\begin{proposition}
For any ordinal $0 < \alpha < \epsilon_0$, there is a unique canonical type $\sigma_\alpha$ with 
$R(\sigma_\alpha)=\alpha$. \QED
\end{proposition}

It is also easy to see by induction on type levels that every uncurried type, and hence every type $\sigma$,
is isomorphic to a unique canonical type (simply by admitting permutations of products 
$\rho_1 \times \cdots \times \rho_n$).
This allows us to extend our ranking $R(-)$ to all types, and we may now define $\sigma \preceq \tau$ iff
$R(\sigma) \leq R(\tau)$.

It is thus clear that $\preceq$ is a total preorder on types, that $\sigma \preceq \tau \preceq \sigma$
iff $R(\sigma) = R(\tau)$ iff $\sigma \sim \tau$, and that $\preceq$ is readily decidable.
We now work towards showing that if $\sigma \preceq \tau$ then $\sigma \lhd_{(\LLL_1,=_1)} \tau$.
This will in fact be easy once we have established a certain way of inductively generating the order relation
on ordinals below $\epsilon_0$.
Let us say a formal sum $\gamma + \alpha$ is a \emph{Cantor sum} if the Cantor normal form of
$\gamma + \alpha$ is $\gamma_1 + \cdots + \gamma_m + \alpha$, 
where $\gamma_1 + \cdots + \gamma_m$ is the Cantor normal form of $\gamma$
(this amounts to the condition that $\alpha \leq \gamma_m$).
Now let $\sqsubseteq$ be the binary relation on ordinals $\alpha < \epsilon_0$ generated by 
the following clauses:
\begin{enumerate}
\item $\alpha \sqsubseteq \alpha$.
\item $\alpha \sqsubseteq \beta \sqsubseteq \gamma$ implies $\alpha \sqsubseteq \gamma$.
\item $\alpha \sqsubseteq \alpha+1$.
\item $\omega^\alpha.k \sqsubseteq \omega^{\alpha+1}$ for any $k < \omega$.
\item If $\alpha \sqsubseteq \beta$ then $\gamma + \alpha \sqsubseteq \gamma + \beta$,
where $\gamma + \alpha$, $\gamma + \beta$ are Cantor sums.
\item If $\alpha \sqsubseteq \beta$ then $\omega^\alpha \sqsubseteq \omega^\beta$.
\end{enumerate}
Clearly if $\alpha \sqsubseteq \beta$ then $\alpha \leq \beta$, since $\leq$ also satisfies
the above properties. Moreover:

\begin{proposition}  \label{ordinal-prop}
If $\alpha \leq \beta < \epsilon_0$ then $\alpha \sqsubseteq \beta$.
\end{proposition}

\noindent \proof :
We show by complete induction on $\beta < \epsilon_0$ that for all $\alpha \leq \beta$ we have
$\alpha \sqsubseteq \beta$.
For $\beta = 0$ this is trivial by clause 1 above.
For the successor case, if the induction claim holds for $\beta$, then for any
$\alpha \leq \beta+1$ we have either $\alpha = \beta+1$, in which case $\alpha \sqsubseteq \beta+1$
by clause 1, or $\alpha \leq \beta$, in which case $\alpha \sqsubseteq \beta \sqsubseteq \beta+1$
by the induction hypothesis and clauses 3 and 2.
For limit ordinals, suppose $\beta$ is expressed as a Cantor sum $\gamma + \omega^\delta$ 
where $\delta > 0$.
If $\delta$ is itself a successor, say $\delta = \zeta+1$, then 
$\beta = \lim_{k < \omega} \gamma + \omega^\zeta.k$, so for any $\alpha \leq \beta$,
either $\alpha = \beta$ (in which case clause 1 applies) or for some $k < \omega$ we have
$\alpha \leq \gamma + \omega^\zeta.k$. But by the induction hypotheses for $\gamma + \omega^\zeta.k$
we have $\alpha \sqsubseteq \gamma + \omega^\zeta.k$, and by clauses 4 and 5 we have
$\gamma + \omega^\zeta.k \sqsubseteq \gamma + \omega^{\zeta+1} = \beta$.
Hence by clause 2 we have $\alpha \sqsubseteq \beta$ as desired.

The remaining case is that $\beta$ is expressed as a Cantor sum $\gamma + \omega^\delta$ 
where $\delta$ is a limit ordinal.
Since $\beta < \epsilon_0$, we have $\delta < \beta$, so we may use the induction hypothesis for $\delta$.
Taking $\delta_0 < \delta_1 < \cdots$ any sequence with limit $\delta$, 
we have that $\beta = \lim_{k < \omega} \gamma + \omega^{\delta_k}$,
so for any $\alpha \leq \beta$ we again have either $\alpha = \beta$ (so that clause 1 applies) or 
$\alpha \leq \gamma + \omega^{\delta_k}$ for some $k$.
But in the latter case, we have $\alpha \sqsubseteq \gamma + \omega^{\delta_k}$
by the induction hypothesis for $\gamma + \omega^{\delta_k}$;
but also $\delta_k \sqsubseteq \delta$ by the induction hypothesis for $\delta$,
whence $\gamma + \omega^{\delta_k} \sqsubseteq \gamma + \omega^\delta = \beta$
by clauses 5 and 6. Hence again $\alpha \sqsubseteq \beta$ by clause 2.
\QED

\vspace*{1.5ex}
The following now establishes the existence of the required retractions.
Note that with Proposition~\ref{ordinal-prop} in hand, 
only the most trivial manipulations of $\lambda$-terms are needed.

\begin{proposition}  \label{retractions-prop}
Whenever $\alpha \leq \beta$, we have $\sigma_\alpha \lhd_{(\LLL_1,=_1)} \sigma_\beta$: 
that is, there are $\LLL_1$ terms $s:\sigma_\alpha \arrow \sigma_\beta$
and $r:\sigma_\beta \arrow \sigma_\alpha$ such that $\lambda x.r(s(x)) =_1 \lambda x.x$.
\end{proposition}

\noindent \proof :
In view of Proposition~\ref{ordinal-prop},
it suffices to show by induction on the generation of $\sqsubseteq$ that
if $\alpha \sqsubseteq \beta$ then $(\LLL_1,=_1) \models \sigma_\alpha \lhd \sigma_\beta$.
This is thus just a question of treating each of the six clauses for $\sqsubseteq$ in turn.
For clauses 1 and 2, we use the usual identity and composition of retractions.
For clause 3, a retraction $\sigma_\alpha \lhd \sigma_{\alpha+1} = \sigma_\alpha \times \nat$
is given by the terms $\lambda x.\ang{x,\num{0}}$ and $\Fst$.
For clause 4, we note that $\sigma_{\omega^\alpha.k} = (\sigma_\alpha \arrow \nat)^k$
(writing $\rho^k$ for the product of $k$ copies of $\rho$) and 
$\sigma_{\omega^{\alpha+1}} = (\sigma_\alpha \times \nat) \arrow \nat$.
We may thus embed the former in the latter by the mapping
\[  \ang{f_1,\ldots,f_k} ~\mapsto~ \lambda \ang{x,z}.~
     \iif_0\,z\,(f_0 x)\;(\iif_1\,z\,(f_1 x)\;(\cdots (\iif_{k-1}\,z\,(f_{k-1} x)\,\num{0}) \cdots )) \]\
and project the latter to the former by the mapping
\[  g ~\mapsto~ \ang{\lambda x.g\ang{x,\num{0}},\cdots,\lambda x.g\ang{x,\num{k-1}}} \, . \]
It is routine to check that the composition of these is $=_1$-convertible to the identity.
For clauses 5 and 6, we use the familiar liftings of a retraction $\sigma_\alpha \lhd \sigma_\beta$
to $\sigma_\gamma \times \sigma_\alpha \lhd \sigma_\gamma \times \sigma_\beta$
and $\sigma_\alpha \arrow \nat \lhd \sigma_\beta \arrow \nat$.
\QED

\begin{theorem}
Whenever $\sigma \preceq \tau$, we have $\sigma \lhd_{(\LLL_1,=_1)} \tau$,
whence $T \models \sigma \lhd \tau$ for any language theory $T \supseteq (\LLL_1,=_1)$.
\end{theorem}

\noindent \proof :
Immediate from Proposition~\ref{retractions-prop} and the trivial isomorphisms
$\sigma \iso \sigma_{R(\sigma)}$, $\tau \iso \sigma_{R(\tau)}$.
\QED

\vspace*{1.5ex}
From the above proofs it is also easy to extract an algorithm which, given any types $\sigma,\tau$
with $\sigma\preceq\tau$, constructs $\LLL_1$ terms $s:\sigma\arrow\tau$ and $r:\tau\arrow\sigma$
that constitute a $(\LLL_1,=_1)$-retraction.

\section{A non-encodability result for pure types}  \label{sec-pure}

It remains to show that if $\sigma \not\preceq \tau$ then no retraction $\sigma \lhd \tau$ can exist
even with respect to $(\PCF,=_\obs)$.
In view of the results of Section~\ref{sec-positive}, it will suffice to show that we never have
$\sigma_{\alpha+1} \lhd_{(\PCF,=_\obs)} \sigma_\alpha$ for any $\alpha<\epsilon_0$:
that is, for no type $\sigma$ can we have $\sigma \times \nat \lhd_{(\PCF,=_\obs)} \sigma$.
In this section we shall establish this for the case when $\sigma$ is a pure type $\pure{k}$
(where $\pure{0} = \nat$ and $\pure{k+1} = \pure{k} \arrow \nat$);
this will introduce many of the key ingredients in a relatively uncluttered form, in preparation for
the general case which we treat in Section~\ref{sec-negative}.

We assume that the reader is familiar with the language $\PCF$ and the associated notion of observational
equivalence, and knows how to set up a version of $\PCF$ with product types and the single base type $\nat$.
We shall write $\PCF^\Omega$ for the extension of $\PCF$ with an `oracle constant' $c_f$ for every (classical) 
partial function $f : \N \parrow \N$.

We also assume familiarity with the \emph{nested sequential procedure} (NSP) model $\SP^0$ for $\PCF$ as
presented in \cite[Chapter~6]{HOC} or \cite{Y-hierarchy}, and with the notation and terminology used
there.
We write $\approx$ for observational equivalence of NSPs, and $\preceq$ for the observational
preorder on them.
As it stands, the model $\SP^0$ does not have product types, but this is not an essential limitation.
Indeed, it is well-known that any type $\sigma$ may be converted to a trivially isomorphic type in 
\emph{curried form} --- that is, one of the form $\sigma_0 \times\cdots\times \sigma_{l-1}$ 
where each $\sigma_i$ is $\times$-free --- in such a way that any $\times$-free $\sigma$ 
is its own curried form.
For a general type $\sigma$ with curried form $\sigma_0 \times\cdots\times \sigma_{l-1}$, 
one may therefore simply define the set $\SP^0(\sigma)$ 
to be the product $\SP^0(\sigma_0) \times\cdots\times \SP^0(\sigma_{l-1})$.

We shall in fact show something a little stronger than the non-existence of a retraction. 
The following concepts will be useful:

\begin{definition}
(i) We say $\sigma$ is a \emph{pseudo-retract} of $\tau$, and write $\sigma \pseudo \tau$, 
if there are closed $\PCF^\Omega$ terms
$T : \sigma \arrow \tau$ and $R : \tau \arrow \sigma$ 
such that $R \circ T \succeq_\obs \id_{\sigma}$, where $\succeq_\obs$ is the
observational preorder on $\PCF$ terms.

Equivalently, in terms of sequential procedures, we may say that if $\sigma,\tau$ respectively have
curried forms $\sigma_0 \times\cdots\times \sigma_{l-1}$ and $\tau_0 \times\cdots\times \tau_{m-1}$,
then a \emph{pseudo-retraction} $\sigma \pseudo \tau$ consists of sequential procedures
\[   z_0^{\sigma_0}, \ldots, z_{l-1}^{\sigma_{l-1}} ~\vdash~ t_0 : \tau_0\,, \ldots, t_{m-1} : \tau_{m-1}\,,
     ~~~~~~
     x_0^{\tau_0}, \ldots, x_{m-1}^{\tau_{m-1}} ~\vdash~ r_0 : \sigma_0\,, \ldots, r_{l-1} : \sigma_{l-1} \]
 such that for each $i<l$ we have $r_i[\vec{x} \mapsto \vec{t}] \succeq z_i^\eta$.
 We say $\sigma$ is a pseudo-retract of $\tau$ if such a pseudo-retraction exists;
 the standard theory of sequential procedures implies that this agrees with the definition via 
 $\PCF^\Omega$ terms.

(ii) A pseudo-retraction $(\vec{t},\vec{r})$ as above is \emph{strict} 
if $\vec{t}\,[\bot_\sigma] \approx \bot_\tau$: more formally, if for all $i<m$ we have
$t_i[\vec{z} \mapsto \vec{\bot}] \approx \bot_{\tau_i}$.

(iii) A pseudo-retraction 
$(\ang{\vec{t},\vec{u}_0,\ldots,\vec{u}_{j-1}},\vec{r}) : \sigma \pseudo \tau \times \rho_0 \times\cdots\times \rho_{j-1}$ 
is \emph{left-strict} with respect to $\tau$ if $\vec{t}\,[\bot_\sigma] \approx \bot_\tau$.
\end{definition}

Although we shall not always bother to distinguish between different ways of bracketing complicated
product types, it is important to note that the concept of left-strictness is defined relative to a certain way
of dividing up the product type on the right-hand side---more specifically, relative to the identification of
$\tau$ as the `first' component of the product. If $\vec\rho$ is empty, then of course left-strictness coincides with strictness.

Our goal in this section will be to prove:

\begin{theorem}  \label{pure-main-thm}
For any $k \geq 0$, the type $\pure{k} \times \nat$ is not a pseudo-retract of $\pure{k}$.
\end{theorem}

This will follow readily from:

\begin{lemma}  \label{pure-main-lemma}
Suppose $k \geq 0$, and $\rho_0,\ldots,\rho_{j-1}$ is any sequence of types of level $<k$.
Then any pseudo-retraction 
$(\ang{t,\vec{u}},r) : \pure{k} \lhd \pure{k} \times \rho_0 \times \cdots \times \rho_{j-1}$ 
must be left-strict with respect to $\pure{k}$.
More formally, given any NSPs
\[  z^k \vdash t : \pure{k} \;, ~~~~~~ z^k \vdash u_i : \rho_i ~~(i<j) \;, ~~~~~~ x^k, \vec{y}^{\,\vec{\rho}} \vdash r : \pure{k} \]
such that $z^k \vdash r[x \mapsto t, \vec{y} \mapsto \vec{u}] \succeq z^\eta$, 
we must have that $t[z \mapsto \bot_k] \approx \bot_k$.
\end{lemma}

We formulate the lemma in terms of a finite sequence of types $\rho_0,\ldots,\rho_{j-1}$ rather than just a single type $\rho$ of level $<k$ so as to cater smoothly for the case $j=0$, when 
$\pure{k} \times \rho_0 \times\cdots\times \rho_{j-1}$ is simply $\pure{k}$.


To see that the lemma implies the theorem, 
suppose we had a pseudo-retraction $\pure{k} \times \nat \pseudo \pure{k}$ comprised by
\[  z':\pure{k},\, y':\nat ~\vdash~ t : \pure{k} \;, ~~~~~~
     x':\pure{k} ~\vdash~ p : \pure{k},~ q : \nat  \]
This gives rise to a pseudo-retraction $\pure{k} \pseudo \pure{k}$ comprised by
\[ z':\pure{k} ~\vdash~ t' \equiv t\,[y \mapsto \lambda.0] : \pure{k} \;, ~~~~~ 
    x':\pure{k} ~\vdash~ p : \pure{k} \]
To see that this is non-strict, we note that
$t'[\bot_k] = t[\bot,\lambda.0] \succeq_\obs t[\bot,\bot]$,
but that $t[\bot,\lambda.0] \not\approx t[\bot,\bot]$ since $q[t[\bot,\lambda.0]] \approx \lambda.0$
whereas $q[t[\bot,\bot]] \preceq q[t[\bot,\lambda.1]] \approx \lambda.1$.
This implies that $t'[\bot_k] \not\approx \bot_k$,
contradicting Lemma~\ref{pure-main-lemma} in the case $j=0$.
%
%
(This argument actually shows that if our language were extended with the unit type $\unit$,
then even $\pure{k} \times \unit$ would not be a retract of $\pure{k}$.)

\vspace*{1.5ex}
The proof of the lemma itself will be modelled largely on the proof of
\cite[Theorem~7.7.1]{HOC} (see also \cite[Theorem~12]{Y-hierarchy} for a slightly improved exposition);
we shall refer to this below as the `standard proof'.
We reason by induction on $k$.

The case $k=0$ is trivial: here we must have $j=0$ since there are no types of level $<0$,
so it suffices to note that if $t[z \mapsto \bot_k] \not\approx \bot_k$
then $z \vdash t \approx \lambda.n$ for some $n$, hence $t$ is not invertible.

Suppose then that $k > 0$ where the lemma holds for $k-1$,
%
and suppose we have
\[  z^k \vdash t : \pure{k} \;, ~~~~~~ z^k \vdash \vec{u} : \vec{\rho} \;, ~~~~~~ 
    x^k, \vec{y}^{\,\vec\rho} \vdash r : \pure{k} \]
where $\lev(\rho) < k$, $z^k \vdash r[x \mapsto t, \vec{y} \mapsto \vec{u}] \succeq z^\eta$.
We wish to show that $t[z \mapsto \bot_k] \approx \bot_k$.

Let \( v \,=\; \dang{r[x \mapsto t, \vec{y} \mapsto \vec{u}]} \),  
so that $z \vdash v \succeq z^\eta$.

\emph{Claim 1:} $v$ has the syntactic form
$\lambda f^{k-1}.\, \caseof{zp}{\cdots}$, where $z^k, f^{k-1} \vdash p : \pure{k-1}$.

\emph{Proof} (transcribed from standard proof):
Clearly $v$ does not have the form 
$\lambda f.n$ or $\lambda f.\bot$, and the only other alternative form
is $\lambda f.\,\caseof{fp'}{\cdots}$. 
In that case, however, we would have 
\[ \dang{v[z \mapsto \lambda w^{k-1}.0]} \cdot\; \bot_{k-1} ~=~ \bot \;, \]
contradicting
$\dang{v[z \mapsto \lambda w^{k-1}.0]} \cdot\; \bot_{k-1} \,\succeq\, (\lambda w.0) \cdot \bot = 0$.
This establishes Claim~1.

Now let $^*$ denote the `dummy substitution' $[z \mapsto \lambda w^{k-1}.0]$.

\emph{Claim 2:} $p^* \succeq f^\eta$, or equivalently $\lambda f.p^* \succeq \id_{k-1}$.

\emph{Proof} (adapted from standard proof):
By the NSP context lemma, it will suffice to show that
$\dang{p^*[f \mapsto s]} \cdot\; q \succeq s \cdot q$ 
for any $s \in \SP^0(k-1)$ and $q \in \SP^0(k-2)$.
(Here and in what follows, the application to $q$ should be omitted in the case $k=1$.)
So suppose $s \cdot q = \lambda.n$ whereas 
$\dang{p^*[f \mapsto s]} \cdot\; q \neq \lambda.n$
for some $n \in \N$.
Take $d = \lambda g.\,\caseof{gq}{n \darrow 0}$,
so that $d \cdot s' = \bot$ whenever $s' \cdot q \neq \lambda.n$.
Then $d \preceq \lambda w.0$ by the context lemma,
so we have $\dang{p[f \mapsto s, z \mapsto d]} \cdot\; q \neq \lambda.n$
since $\lambda.n$ is maximal in $\SP^0(\nat)$.
By the definition of $d$, it follows that
$\dang{(zp)[f \mapsto s, z \mapsto d]}\, = \bot$, whence
$\dang{v[z \mapsto d]} \cdot\, s = \bot$, whereas $d \cdot s=0$,
contradicting $\dang{v[z \mapsto d]}\, \succeq d$.
This completes the proof of Claim~2.


Now consider the head reduction sequence
\[ z ~\vdash~ r[x \mapsto t, \vec{y} \mapsto \vec{u}] ~~\reducesto_h^*~~ \lambda f^{k-1}.\,\caseof{zP}{\cdots} \;, \]
where $\dang{\!P\!} \,= p$. The head $z$ on the right hand side will have some ancestor within
$r[x \mapsto t, \vec y \mapsto \vec u]$; and since $z$ is not free in $r$, this must appear as the head of
some application $zp'$ within either $\vec u$ or $t$.

\emph{Case 1:} $zp'$ comes from $\vec u$, say from $u_j$. Since $z^k \vdash u_j : \rho_j$ 
where $\lev(\rho_j) \leq k-1$,
all bound variables within $u_j$ are of level $\leq k-2$. Let $\vec{y'}$ be the list of bound variables of $u_j$
in scope at the critical occurrence of $zp'$.
Then as in the standard proof, by tracking the subterm $zp'$ through the above reduction sequence,
we easily see that $z,f \vdash P = p' [\vec{y'} \mapsto \vec{T'}]$ for some meta-terms $z, f \vdash \vec{T'}$.
So we have
\[ f^{k-1} ~\vdash~ p'^*[\vec{y'} \mapsto \vec{T'}^*] ~=~ P^* ~\approx~ p^* ~\succeq~ f^\eta \;. \]
This exhibits $\pure{k-1}$ as a retract of some finite product of level $\leq k-2$ types 
$\rho'_0,\ldots,\rho'_{j'-1}$, contradicting Theorem~7.7.1 of \cite{HOC}. 
Alternatively, it contradicts the induction hypothesis of the present proof,
since we can easily extend this to a retraction 
$(\ang{t',u'},r'): \pure{k-1} \lhd \pure{k-1} \times \rho'_0 \times\cdots\times \rho'_{j'-1}$
where $f \vdash t' = \lambda w'.0$ (or just $\lambda.0$ in the case $k=1$).

\emph{Case 2:} $zp'$ comes from $t$. 
Write $t'$ as $\lambda x'.e'$,
where $x'$ has type $\pure{k-1}$, but all variables bound within $e'$ are of level $\leq k-2$.
Let $\vec{y'}$ be the list of bound variables of $e'$ that are in scope at the critical occurrence of $zp'$.
Then as above, we have that
$z,f \vdash P = p' [x' \mapsto T', \vec{y'} \mapsto \vec{U'}]$ for some meta-terms $z,f \vdash T',\vec{U'}$.
So we have
\[ f^{k-1} ~\vdash~ {p}'^*[x' \mapsto {T'}^*, \vec{y'} \mapsto \vec{U'}^*] ~=~ P^* ~\approx~ p^* 
   ~\succeq~ f^\eta \;. \]
Let $\rho'$ denote the product of the types of the $\vec{y'}$.
Then $\lev(\rho') \leq k-2$, and the above constitutes a pseudo-retraction
$(\ang{t',\vec{u'}},r') : \pure{k-1} \lhd \pure{k-1} \times \rho'$,
where $t'$ is given by ${T'}^*$, $\vec{u'}$ by $\vec{U'}^*$, and $r'$ by ${p'}^*$.

By the induction hypothesis, this pseudo-retraction is left-strict w.r.t.\ $\pure{k-1}$:
that is, $t'[\bot_{k-1}] \approx \bot_{k-1}$, or more formally 
${T'}^*[f \mapsto \bot_{k-1}] \approx \bot_{k-1}$.
To show that this implies that our original pseudo-retraction $(\ang{t,u},r)$ is left-strict,
we require a further argument that did not feature in the standard proof.

\emph{Claim 3:} $t$ has head variable $z$, whence $t[z \mapsto \bot_k] \approx \bot_k$.

\emph{Proof:} Since $t$ somewhere contains the application $zp'$, $t$ is not a constant procedure,
and the only other alternative is that $t$ has the form $\lambda x'.\,\caseof{x'q}{\cdots}$ 
(omitting $q$ in the case $k=1$).
By tracking the transformation of $zp'$ to $zP$ through the head reduction sequence for
$r[x \mapsto t, \vec y \mapsto \vec u]$, we now see that this sequence must have contained reductions
\begin{eqnarray*}
\lambda f.\,\caseof{t_0 T'}{\cdots} & \equiv & \lambda f.\,\caseof{(\lambda x'.\caseof{x'q}{\cdots})T'}{\cdots} \\
  & \reducesto_h & \lambda f.\,\caseof{(\caseof{T'q'}{\cdots})}{\cdots} \\
  & \reducesto_h & \lambda f.\,\caseof{T'q'}{\cdots} \;, 
\end{eqnarray*}
where $q' = q[x' \mapsto T']$. Specializing via $^*$, we obtain
\[ r[x \mapsto t^*, \vec y \mapsto \vec u^*] ~~\reducesto_h^*~~ \lambda f.\,\caseof{{T'}^*{q'}^*}{\cdots} \]
where $f \vdash {T'}^* {q'}^*$.
But $r[x \mapsto t^*, \vec y \mapsto \vec u^*] \succeq {z^\eta}^* \approx \lambda w.0$,
so $\dang{r[x \mapsto t^*, \vec y \mapsto \vec u^*]}$ can only be the procedure $\lambda f.0$.
Thus the subterm $\caseof{{T'}^*{q'}^*}{\cdots}$ above evaluates to $0$, and so 
${T'}^*{q'}^*$ itself must evaluate to some numeral, say $m$.
Finally, specializing $f$ to $\bot$, we have 
$({T'}^*[f \mapsto \bot])({q'}^*[f \mapsto \bot]) \reducesto^* m$,
so ${T'}^*[f \mapsto \bot] \not\approx \bot$, contrary to what was established above
by the induction hypothesis.
(For the case $k=1$, the references to $q,q',{q'}^*$ should of course be deleted.)

The second part of the claim follows trivially, giving the desired left-strictness of $(\ang{t,u},r)$.
This completes the proof of Lemma~\ref{pure-main-lemma}, 
and hence of Theorem~\ref{pure-main-thm}.

\section{Non-encodability in the general case}  \label{sec-negative}

We now wish to generalize the above proof to show that:

\begin{theorem}  \label{main-theorem}
The type $\sigma \times \nat$ is not a pseudo-retract of $\sigma$ for any $\sigma$.
\end{theorem}

This will follow from a lemma proved by induction on the ordinal rank of $\sigma$.
In the general setting, however, this lemma will need to be formulated somewhat more subtly than
Lemma~\ref{pure-main-lemma}.
Motivated by the arguments of the previous section, we introduce the following concept:

\begin{definition}  \label{quasi-retraction-def}
Suppose $\vec{\tau}$ and $\vec{\rho}$ are lists of $\times$-free types.
A \emph{quasi-retraction} $\vec{\tau} \isplit \vec{\rho}$ consists of terms
\[ z : \vec{\tau} \arrow \nat ~\vdash~ u : \vec{\rho} \arrow \nat \;, ~~~~~~
   x : \vec{\rho} \arrow \nat ~\vdash~ r : \vec{\tau} \arrow \nat \] 
such that we have a head reduction sequence
\[ z ~\vdash~ r[x \mapsto u] ~~\reducesto_h^*~~
   \lambda \vec{f}^{\,\vec{\tau}}.\;\caseof{z\vec{P}}{\cdots}  \]
where $\vec{P}[z \mapsto \lambda \vec{w}.0] \succeq \vec{f}^{\,\eta}$.
\end{definition}

The reader will see that a situation very close to this appeared in the course of the proof of
Lemma~\ref{pure-main-lemma}.
As we will shortly see, a quasi-retraction $\vec\tau \isplit \vec\rho$ gives rise to a pseudo-retraction
$\vec\tau \pseudo \vec\rho \times \vec\upsilon$ where the $\vec \upsilon$ are in some sense `lower' types.
Nevertheless, in the general setting, the existence of a quasi-retraction turns out to afford a more
suitable induction hypothesis than the existence of such a pseudo-retraction, since the former implicitly 
imposes some useful additional constraints on how these $\vec\upsilon$ components will behave.

We shall often identify a list of types $\vec{\tau} = \tau_0,\ldots,\tau_{h-1}$ with the corresponding
product type $\tau_0 \times\cdots\times \tau_{h-1}$. Thus, if $\tau$ and $\rho$ are any types in curried
form, respectively the products of the lists $\vec{\tau}$ and $\vec{\rho}$, then we may refer to 
a quasi-retraction $(u,r) : \vec{\tau} \isplit \vec{\rho}$ also as 
a quasi-retraction $\tau \isplit \rho$.

In the situation of Definition~\ref{quasi-retraction-def}, 
the head $z$ on the right-hand side will originate from some application subterm $z \vec{p}$ within $u$. 
Let $\vec{y}$ be the list of bound variables of $u$ in scope at the point of this subterm's occurrence.
These will consist of the top-level bound variables of $u$, of types $\vec{\rho}$ (we call these the
\emph{major} variables of $\vec{y}$),
plus possibly some others, say of types $\vec{\upsilon}$ (we call these the \emph{minor} variables).
The latter will be associated with applications $y_i \vec{q}$
within $u$ that contain the critical occurrence of $z$. (Note that none are associated with applications
$z \vec{q}$, since the outer $z$ would then prevent the critical $z$ from emerging as the head variable.)
It follows that if the types $\vec\rho$ are all of level $\leq k$, then the types $\vec\upsilon$
are all of level $\leq k-2$; however, there may be types among $\vec\upsilon$ that are higher than
some among $\vec\rho$.

We also have in this situation that $z,\vec{f} \vdash \vec{P} \equiv \vec{p}\,[\vec{y} \mapsto \vec{T}]$
for some meta-terms $\vec{T}$.
Writing $^*$ for the substitution $[z \mapsto \lambda \vec{w}.0]$, it follows that
$\vec{f} \vdash \vec{p}\,^*[\vec{y} \mapsto \vec{T}^*] \equiv \vec{P}^* \succeq \vec{f}^{\,\eta}$,
so that $(\vec{T}^*,\vec{p}\,^*)$ exhibit a pseudo-retraction 
$\vec{\tau} \pseudo \vec{\rho} \times \vec{\upsilon}$.
We call this the \emph{associated pseudo-retraction} of the quasi-retraction $(u,r)$,
and refer to the $\vec{\rho}$ and $\vec{\upsilon}$ as its \emph{major} and \emph{minor} components respectively.

\begin{definition}
We say a quasi-retraction $\vec{\tau} \,\isplit\, \sigma \times \vec{\rho}$
is \emph{left-strict} with respect to $\sigma$ if the associated pseudo-retraction
$\vec{\tau} \pseudo \sigma \times \vec{\rho} \times \vec{\upsilon}$ is left-strict with respect to $\sigma$.
In the case that $\vec{\rho}$ is empty, we may also say simply that such a quasi-retraction
$\vec{\tau} \isplit \sigma$ is \emph{strict}.
\end{definition}
Once again, we note that the notion of a left-strict quasi-retraction is defined relative to some identification
of the `first' component of the product on the right hand side.

It will be helpful to know that any quasi-retraction can be replaced by an `equivalent' one of a more
restricted form. Specifically, we shall say a quasi-retraction 
$(z \vdash u,\, x \vdash r) : \vec\tau \isplit \vec\rho$ is \emph{simple}
if $r$ contains just a single free occurrence of $x$, and this is at the head of $r$
(i.e.\ $r$ has the form $\lambda \vec{f}.\,\caseof{x\vec{q}}{\cdots}$).
We then have:

\begin{proposition}  \label{simple-qr-prop}
Given any quasi-retraction $(u,r): \vec{\tau} \isplit \vec{\rho}$,
there is a simple quasi-retraction $(u,r_1): \vec{\tau} \isplit \vec{\rho}$
with the same associated pseudo-retraction as $(u,r)$.
In particular, if $\vec{\rho} = \vec{\sigma},\vec{\rho'}$, then
$(u,r_1)$ is left-strict w.r.t.\ $\vec{\sigma}$ iff $(u,r)$ is.
\end{proposition}

\noindent \proof :
Suppose we are given $(z \vdash u,\, x \vdash r): \vec{\tau} \isplit \vec{\rho}$.
Perform head reductions on $r [x \mapsto u]$, tracking residuals of the substituted occurrences of $u$,
until the occurrence of $u$ containing the critical $z\vec{p}$ appears in head operator position.
Let $r_0$ be obtained by replacing this occurrence of $u$ (only) by $x$, so that we have
\[ r[x \mapsto u] ~~\reducesto_h^*~~ r_0[x \mapsto u]
   ~~\reducesto_h^*~~ \caseof{z P_0 \ldots P_{k-1} \vec{Q}}{\cdots} \;. \]
This need not yield a quasi-retraction as it stands, since $r_0$ may contain free occurrences of 
$z$ within other occurrences of $u$,
although these will never come into head position in the course of head reduction.
Setting $r_1 = r_0 [z \mapsto \lambda \vec{w}.0]$, the second half of the above reduction thus
specializes to 
\[  r_1 [x \mapsto u] ~~\reducesto_h^*~~ \caseof{z \vec{P_1} \vec{Q_1}}{\cdots}  \]
where $\vec{P_1},\vec{Q_1}$ are obtained from $\vec{P},\vec{Q}$ by possibly replacing 
certain occurrences of $z$ by $\lambda \vec{w}.0$.
We therefore have $\vec{P_1}^{\,*} = \vec{P}^{\,*} \succeq \vec{f}^{\,\eta}$ and likewise
$\vec{Q_1}^{\,*} \succeq \vec{g}^{\,\eta}$, so $u,r_1$ constitute a quasi-retraction with the
same associated pseudo-retraction as $u,r$, and by construction of $r_1$ it is clear that this is simple.

The last clause in the proposition is immediate, as the left-strictness of a quasi-retraction is determined
by the associated pseudo-retraction.
\QED

\vspace*{1.5ex}
One could in fact go further and show that any quasi-retraction may replaced by a simple one for which
the resulting meta-terms $\vec{P}$ do not contain $z$ free.
This may seem an aesthetic improvement in that it eliminates the need for the `arbitrary' specialization
$z \mapsto \lambda\vec{w}.0$, but it appears not to lead to any actual simplification in our main proof.

We may now formulate our main lemma as follows.
If $\alpha,\gamma < \epsilon_0$, we shall write $\alpha \gg \gamma$ if
$\alpha_{k-1} > \gamma$, where $\alpha_0 +\cdots+ \alpha_{k-1}$ is the Cantor normal form of $\alpha$.

\begin{lemma}  \label{main-lemma}
If $R(\sigma') \leq R(\sigma)$ and $R(\sigma) \gg R(\vec\rho)$, then
any quasi-retraction $\sigma \isplit \sigma' \times \vec{\rho}$ must be left-strict with respect to $\sigma'$.
\end{lemma}

Before proving this, we pause to show:


\begin{proposition}   \label{lemma-implies-theorem}
Lemma~\ref{main-lemma} implies Theorem~\ref{main-theorem}.
\end{proposition}

\noindent \proof :
Suppose we had a pseudo-retraction $\sigma \times \nat \pseudo \sigma$ comprised by
\[  z':\sigma,\, y':\nat ~\vdash~ t : \sigma \;, ~~~~~~
     x':\sigma ~\vdash~ p : \sigma,~ q : \nat  \]
This gives rise to a pseudo-retraction $\sigma \pseudo \sigma$ comprised by
\[ z':\sigma ~\vdash~ t' \equiv t\,[y \mapsto \lambda.0] : \sigma \;, ~~~~~ x':\sigma ~\vdash~ p : \sigma \]
which can be seen to be non-strict just as in the corresponding argument in Section~\ref{sec-pure}.


We may build a quasi-retraction $(u,r) : \sigma \isplit \sigma$ 
with $(t',p)$ as its associated pseudo-retraction (the minor component being empty): take 
\[ z : \sigma \arrow \nat ~\vdash~ u \equiv \lambda x'.\, \caseof{z p}{i \darrow i} \;, ~~~~~~
   x : \sigma \arrow \nat ~\vdash~ \lambda z'. e \equiv \lambda \vec{f}.\, \caseof{x t'}{i \darrow i}  \;, \]
so that 
$z \vdash r[x \mapsto u] \reducesto_h^* \lambda z'.\,\caseof{z (p[x' \mapsto t'])}{\cdots}$ 
where $p[x' \mapsto t'][z \mapsto \lambda \vec{w}.0] \succeq z'^{\,\eta}$.
We thus have a non-strict quasi-retraction $\sigma \isplit \sigma$, 
contradicting Lemma~\ref{main-lemma}.
\QED

\vspace*{1.5ex}
In a similar vein, it is not hard to deduce from Lemma~\ref{main-lemma} 
that in the situation of the lemma we must actually have $R(\sigma')=R(\sigma)$.
We may infer this from:

\begin{proposition}  \label{ranks-match-prop}
From any quasi-retraction $\vec{\tau} \isplit \vec{\sigma} \times \vec{\rho}$ we may obtain a 
quasi-retraction $\vec{\tau} \isplit (\vec{\sigma} \times \nat) \times \vec{\rho}$
which is not left-strict with respect to $(\vec{\sigma} \times \nat)$.
\end{proposition}

\noindent \proof :
Suppose $\vec{\tau} \isplit \vec{\sigma} \times \vec{\rho}$ is witnessed by terms
\[ z : \vec{\tau} \arrow \nat ~\vdash~ 
        u \equiv \lambda \vec{a}^{\,\vec{\sigma}} \vec{c}^{\,\vec{\rho}}.\,d ~:~ 
        \sigma \arrow \vec{\rho} \arrow \nat \,, ~~~~~~ 
   x : \vec{\sigma} \arrow \vec{\rho} \arrow \nat ~\vdash~ r : \vec{\tau} \arrow \nat \;. \]
Define $u' \equiv \lambda \vec{a}^{\,\vec{\sigma}} b^\nat \vec{c}^{\,\vec{\rho}}.\,d$,
so that $z \vdash u'  : \vec{\sigma} \arrow \nat \arrow \vec{\rho} \arrow \nat$.
Now take a new variable $x : \vec{\sigma} \arrow \nat \arrow \vec{\rho} \arrow \nat$,
and obtain $r'$ from $r$ by replacing all applications $x \vec{p} \vec{q}$ by $x' \vec{p} (\lambda.0) \vec{q}$,
so that $x' \vdash r' : \vec{\tau} \arrow \nat$.
It is then clear that $r'[x \mapsto u'] \approx r[x \mapsto u]$, so these two terms head-reduce to
$\approx$-equivalent head normal forms, whence $u',r'$ constitute a quasi-retraction
$\vec{\tau} \isplit \sigma \times \nat \times \vec{\rho}$.
However, this quasi-retraction is not left-strict w.r.t.\ $\sigma\times\nat$:
the new variable $b^\nat$ will be specialized in the course of the head reduction to $\lambda.0$,
so the induced mapping $\vec{\tau} \arrow (\vec{\sigma}\times\nat)$ will be everywhere non-$\bot$
in the second component.
\QED

\vspace*{1.5ex}
Applying this to the situation of Lemma~\ref{main-lemma}, we see that if we had
$\sigma \isplit \sigma' \times \vec{\rho}$ where $R(\sigma') < R(\sigma) \gg R(\vec{\rho})$,
then Proposition~\ref{ranks-match-prop} would yield a non-left-strict
$\sigma \isplit (\sigma' \times \nat) \times \vec{\rho}$ where $R(\sigma'\times\nat) \leq R(\sigma)$,
contrary to the lemma itself.


\vspace*{1.5ex}
We now proceed to the proof of Lemma~\ref{main-lemma}.
Without loss of generality we may work with types $\sigma = \sigma_\alpha$ and 
$\sigma' = \sigma_{\alpha'}$, where $\alpha,\alpha' < \epsilon_0$ and $\alpha \gg R(\vec{\rho})$;
we require to show that any $\sigma \isplit \sigma' \times \vec{\rho}$ is left-strict w.r.t.\ $\sigma'$.
We reason by transfinite induction on $\alpha$.

\vspace*{1.5ex}
\emph{Case 1}: $\alpha$ is a finite ordinal $k>0$, so that $\sigma = \sigma_\alpha = \nat^k$.
In this case $\vec\rho$ must be empty, since there are no types of lower rank than $\nat$.
Moreover, in the light of Proposition~\ref{ranks-match-prop} it will suffice to treat the case 
$\sigma' = \nat^k$.

Any quasi-retraction $\nat^k \isplit \nat^k$ would have an associated pseudo-retraction
$\nat^k \pseudo \nat^k$, since in this case the minor component must be empty.
Moreover, we claim that any such pseudo-retraction must actually be a retraction,
in that if $f : \nat^k \arrow \nat^k$ and
$f \succeq_\obs \id_{\nat^k}$ then actually $f \approx_\obs \id$. 
For otherwise, we could take $x \in \SP(\nat^k)$ with $f \cdot x \succeq_\obs x$ but $f \cdot x \neq x$,
and then pick $y \succeq_\obs x$ so that $y$ and $f \cdot x$ had no upper bound.
This gives a contradiction since in fact $f \cdot y \succeq y$ and $f \cdot y \succeq x$.

Finally, we claim that any retraction $(t,r) : \nat^k \lhd \nat^k$ is strict, i.e.\ that
$t[\vec{\bot}] = \vec{\bot}$. To see this, we note that $\nat^k$ contains a strictly ascending chain
of length $k+1$ with least element $\vec{\bot}$, and this must be mapped by $t$ to strictly ascending
chain of length $k+1$ with least element $t[\vec{\bot}]$. But this means that $t[\vec{\bot}] = \vec{\bot}$,
since $\nat^k$ contains no strictly ascending chains of length $k+2$.
We have thus shown that any quasi-retraction $\nat^k \isplit \nat^k$ must be strict.

\vspace*{1.5ex}
\emph{Case 2}: $\alpha = \omega^\delta.k$ for some $\delta>0$ and $k \in \N$.
The argument here will be an elaboration of the proof of Lemma~\ref{pure-main-lemma}.
We write $\sigma = \sigma_\alpha$ in curried form as 
$(\tau_0 \arrow\cdots\arrow \tau_{h-1} \arrow \N)^k$
or $(\vec{\tau}\arrow\nat)^k$ (where $\vec{\tau}$ may be empty).
We also write $\sigma'$ in curried form as $\sigma'_0 \times\cdots\times \sigma'_{l-1}$, 
where each $\sigma'_i$ has rank $\leq \omega^\delta$.

Suppose then that we have a quasi-retraction 
$\sigma \isplit \sigma' \times \vec{\rho}$, 
where $R(\sigma') \leq \omega^\delta.k$ and $R(\vec\rho) < \omega^\delta$.
This will have an associated pseudo-retraction 
$\sigma \pseudo \sigma' \times (\vec{\rho} \times \vec{\upsilon})$,
where also $R(\vec\upsilon) < \omega^\delta$. 
Since $\vec{\rho}$ and $\vec{\upsilon}$ here have the same status,
we henceforth amalgamate them and call them $\vec{\rho^+}$, 
which we may further assume to be
a sequence of $\times$-free types of non-increasing rank.
We will in fact show that \emph{any} pseudo-retraction $\sigma \pseudo \sigma' \times \vec{\rho^+}$
is left-strict w.r.t.\ $\sigma'$.

At the level of NSPs, such a pseudo-retraction amounts to having procedures
\[  z_0,\ldots,z_{k-1} : \vec{\tau}\arrow\nat ~\vdash~ 
    t_0 : \sigma'_0, ~ \ldots,~ t_{l-1} : \sigma'_{l-1}, ~ 
     \vec{u} : \rho^+ \;, \]
\[    x_0:\sigma'_0,~ \ldots,~ x_{l-1}:\sigma'_{l-1},~ \vec{y}:\vec{\rho^+} 
      ~\vdash~ r_0,\ldots,r_{k-1} : \vec{\tau}\arrow\nat \]
such that $\lev(\rho) < \omega^\delta$, 
$\vec{z} \vdash r[\vec{x} \mapsto \vec{t}, \vec{y} \mapsto \vec{u}\,] \succeq z_i^\eta$ for each $i$.
Let \[ v_i ~=~ \dang{r_i[\vec{x} \mapsto \vec{t}, \vec{y} \mapsto \vec{u}\,]} \] 
for each $i<k$, so that $\vec{z} \vdash v_i \succeq z_i^\eta : \vec{\tau} \arrow \nat$.

\emph{Claim 1:} Each $v_i$ has the syntactic form
$\lambda \vec{f}^{\,\vec{\tau}}.\, \caseof{z_i p_{i0} \ldots p_{i(h-1)}}{\cdots}$, 
where $z_i^{\vec{\tau}\arrow\nat}, \vec{f}^{\,\vec{\tau}} \vdash p_{ij} : \tau_j$ for each $j<h$.

\emph{Proof}:
Clearly $v_i$ does not have the form $\lambda \vec{f}.n$ or $\lambda \vec{f}.\bot$, 
nor the form $\lambda \vec{f}.\,\caseof{z_j \vec{p'}}{\cdots}$ for $j\neq i$, 
since this would contradict $v_i \succeq z_i^\eta$ if we specialized $z_j$ to $\bot$
and $z_i$ to $\lambda \vec{w}.0$.
The only other alternative form for $v_i$ is $\lambda \vec{f}.\,\caseof{f_j \vec{q}}{\cdots}$
for some $j < h$. 
In that case, however, we would have 
\[ \dang{v[z_0,\ldots,z_{k-1} \mapsto \lambda \vec{w}^{\,\vec{\tau}}.0]} \cdot\; \vec{\bot}_{\vec{\tau}} 
   ~=~ \bot_\nat \;, \]
contradicting
\( \dang{v[\vec{z} \mapsto \lambda \vec{w}.0]} \cdot\; \vec{\bot}_{\vec{\tau}} 
   \,\succeq\, (\lambda \vec{w}.0) \cdot \vec{\bot}_{\vec{\tau}} =0 \).
This establishes Claim~1.

We henceforth write $^*$ for the substitution 
$[z_0,\ldots,z_{k-1} \mapsto \lambda \vec{w}^{\,\vec{\tau}}.0]$.
We now look more closely at the subterms $p_{ij}$ appearing in Claim~1.

\emph{Claim 2:} For each $i,j$, we have $\vec{z},\vec{f} \vdash p_{ij}^* \succeq f_j^\eta$. 

\emph{Proof}:
By the NSP context lemma, it will suffice to show that
$\dang{p_{ij}^*[\vec{f} \mapsto \vec{s}\,]} \cdot\; \vec{q} \succeq s_j \cdot \vec{q}$ 
for any $\vec{s} \in \SP^0(\vec{\tau})$ and any closed $\vec{q}$ of length and types appropriate to $\tau_j$.
So suppose for contradiction that for some $n \in \N$ we have $s_j \cdot \vec{q} = \lambda.n$ 
whereas $\dang{p_{ij}^*[\vec{f} \mapsto \vec{s}\,]} \cdot\; \vec{q} \neq \lambda.n$.
Take $d = \lambda \vec{g}.\,\caseof{g_j\vec{q}}{n \darrow 0}$,
so that $d \cdot \vec{s} = \lambda.0$,
but $d \cdot \vec{s'} = \bot$ whenever $s'_j \cdot \vec{q} \neq \lambda.n$.
Then $d \preceq \lambda \vec{w}.0$ by the context lemma,
so we have 
\[ \dang{p_{ij}[\vec{f} \mapsto \vec{s}, \vec{z} \mapsto d]} \cdot\; \vec{q} ~\neq~ \lambda.n \]
since otherwise 
$\dang{p_{ij}[\vec{f} \mapsto \vec{s}, \vec{z} \mapsto \lambda \vec{w}.0]} \cdot\; \vec{q} = \lambda.n$.
It follows from the aforementioned property of $d$ that
\[ \dang{(z_i \vec{p_i})[\vec{f} \mapsto \vec{s}, \vec{z} \mapsto d]}
   ~=~ d \;\cdot \dang{\vec{p_i}\,[\vec{f} \mapsto \vec{s}, \vec{z} \mapsto d]} ~=~ \bot \;, \]
whence $\dang{v_i[\vec{z} \mapsto d]} \cdot\, \vec{s} = \bot$, whereas $d \cdot \vec{s} = \lambda.0$.
But this contradicts 
$\dang{v_i[\vec{z} \mapsto d]}\;\succeq\; \dang{z_i^\eta[\vec{z} \mapsto d]}\, \approx d$.
This completes the proof of Claim~2.


Next, we note that in view of Claim~1, for each $i<k$ we will have some head reduction sequence
\[ \vec{z} ~\vdash~ r_i[\vec{x} \mapsto \vec{t}, \vec{y} \mapsto \vec{u}\,] ~~\reducesto_h^*~~ 
   \lambda \vec{f}^{\,\vec{\tau}}.\,\caseof{z_i P_{i0} \ldots P_{i(h-1)}}{\cdots} \;, \]
where $\dang{\!P_{ij}\!} \,= p_{ij}$ for each $i,j$. 
In each case, the head $z_i$ on the right hand side will have some ancestor within
$r[\vec{x} \mapsto \vec{t}, \vec{y} \mapsto \vec{u}\,]$; and since the $z_i$ are not free in $r$, 
this must appear as the head of some application $z_i \vec{p'_i}$ appearing in either $u$ or $t$.
We now consider various possible situations in turn:

\vspace*{1.0ex}
\emph{Subcase 2.1:} For some $i$, the critical application $z_i \vec{p'_i}$ comes from some $u_j$. 
Recall that $\vec{z} \vdash u_j : \rho^+_j$ where $R(\rho^+_j) \leq \omega^\delta$.
Since $\rho^+_j$ is $\times$-free, we have $R(\rho^+_j) = \omega^\gamma$ for some $\gamma < \delta$,
and we may write $\rho^+_j$ as $ \vec{\upsilon'} \arrow \nat$, 
where $\vec{\upsilon'}$ is a sequence of non-increasing rank.

We may specialize the above head reduction sequence to one that exhibits a quasi-retraction 
$\vec{\tau} \isplit \vec{\upsilon'}$ as follows.
Let $\vec{y}^{\,\circ}$, $\vec{u}^{\,\circ}$ denote the lists $\vec{y}, \vec{u}$ with (respectively)
$y_j$ and $u_j$ deleted, and let $\dag$ denote the substitution mapping the variables 
$\vec{z}$ except for $z_i$ to $\lambda \vec{w}.0$.
Let $u' = u_j^\dag$
and let $r' = (r_i[\vec{x} \mapsto \vec{t}, \vec{y}^{\,\circ} \mapsto \vec{u}^{\,\circ}])^\dag$,
so that $z_i \vdash u' : \rho^+_j$ , $y_j \vdash r' : \vec{\tau} \arrow \nat$, 
and $r'[y_j \mapsto u'] \equiv (r_i[\vec{x} \mapsto \vec{t}, \vec{y} \mapsto \vec{u}\,])^\dag$.
Then the above head reduction specializes to a head reduction
\[ z_i ~\vdash~ r'[y_j \mapsto u'] ~~\reducesto_h^*~~ 
    \lambda \vec{f}^\tau.\;\caseof{z_i P'_0 \ldots P'_{h-1}}{\cdots} \;, \]
where $P'_j = P_{ij}^\dag$ and so $P'_j[z_i \mapsto \lambda \vec{w}.0] = P_{ij}^* \succeq f_j^\eta$
for each $j$. This gives $\vec{\tau} \isplit \vec{\upsilon'}$.
We may now apply Proposition~\ref{ranks-match-prop} (with $\vec{\rho}$ empty)
to obtain a non-strict quasi-retraction $\vec{\tau} \isplit \vec{\upsilon'} \times \nat$.
Since $R(\vec{\upsilon'} \times \nat) \leq R(\vec{\tau})$,
this contradicts our induction hypothesis for $\delta$.

\vspace*{1.0ex}
\emph{Subcase 2.2:} For every $i<k$, the critical application $z_i \vec{p'_i}$ comes from some
$t_{j(i)}$ headed by one of the variables $\vec{z}$.
Then in each case, the head variable of $t_{j(i)}$ is necessarily $z_i$ itself, otherwise $z_i$ would
not emerge as the head variable in the head reduction above.
This means that $t_{j(i)}, t_{j(i')}$ are distinct whenever $i \neq i'$.
Thus every one of $t_0,\ldots,t_{k-1}$ arises as $t_{j(i)}$ for some $i$, so that they all have
head variables drawn from $\vec{z}$.
It follows that $\vec{t}[\vec{z} \mapsto \vec{\bot}] \approx \vec{\bot}$,
which is to say that our pseudo-retraction $(\ang{\vec{t},u},\vec{r})$ is left-strict as required.


\vspace*{1.0ex}
\emph{Subcase 2.3:} 
Some critical application $z_i \vec{p'_i}$ comes from a $t_c$ not headed by one of the $\vec{z}$.
Clearly $t_c$ is not a constant procedure $\lambda \vec{x'}.m$ for $m \in \N_\bot$, since this contains
no application of $z_i$. We must therefore have that
$t_c = \lambda \vec{x'}.\,\caseof{x'_j \vec{q}}{\cdots}$ for some $j$,
where the variables $\vec{x'}$ have certain types $\vec{\tau'}$ with $R(\vec{\tau'}) \leq R(\vec{\tau})$.

We may then specialize the $i$th head reduction above to one that exhibits a quasi-retraction 
$\vec{\tau} \isplit \vec{\tau'}$:
let $\vec{x}^{\,\circ}$, $\vec{t}^{\,\circ}$ denote the lists $\vec{x}, \vec{t}$ with (respectively)
$x_c$ and $t_c$ deleted, and again let $\dag$ denote the substitution mapping the variables 
$\vec{z}$ except for $z_i$ to $\lambda \vec{w}.0$.
Let $t' = t_c^\dag$, 
and let $r' = (r_i[\vec{x}^{\,\circ} \mapsto \vec{t}^{\,\circ}, \vec{y} \mapsto \vec{u}])^\dag$,
so that $z_i \vdash t' : \vec{\tau'}\arrow\nat$ , $x_c \vdash r' : \vec{\tau} \arrow \nat$, 
and $r'[x_c \mapsto t'] \equiv (r_i[\vec{x} \mapsto \vec{t}, \vec{y} \mapsto \vec{u}])^\dag$.
Then the above head reduction specializes to a head reduction
\[ z_i ~\vdash~ r'[x_c \mapsto t'] ~~\reducesto_h^*~~ 
    \lambda \vec{f}^\tau.\;\caseof{z_i P'_0 \ldots P'_{h-1}}{\cdots} \;, \]
where $P'_j = P_{ij}^\dag$ and so $P'_j[z_i \mapsto \lambda \vec{w}.0] = P_{ij}^* \succeq f_j^\eta$
for each $j$. This constitutes a quasi-retraction $\vec{\tau} \isplit \vec{\tau'}$.


Since $R(\vec{\tau'}) \leq R(\vec{\tau})$, our induction hypothesis for $\delta$ tells us that this
quasi-retraction must be strict. But this leads to a contradiction in view of the following argument.

\emph{Claim 3:} The above quasi-retraction $\vec\tau \isplit \vec\tau$ is non-strict.


\emph{Proof:}
By tracking the transformation of $z_i \vec{p'_i}$ to $z_i \vec{P_i}$ through the head reduction sequence for
$r_i[\vec{x} \mapsto \vec{t}, y \mapsto u]$, we see that this sequence must have contained reductions
\begin{eqnarray*}
\lambda \vec{f}.\,\caseof{t_c \vec{T'}}{\cdots} 
  & \equiv & \lambda \vec{f}.\,\caseof{(\lambda \vec{x'}.\,\caseof{x'_j \vec{q}}{\cdots})\vec{T'}}{\cdots} \\
  & \reducesto_h & \lambda f.\,\caseof{(\caseof{T'_j \vec{q'}}{\cdots})}{\cdots} \\
  & \reducesto_h & \lambda f.\,\caseof{T'_j \vec{q'}}{\cdots} \;, 
\end{eqnarray*}
where $\vec{q'} = \vec{q}\,[\vec{x'} \mapsto \vec{T'}]$. Specializing via $^*$, we obtain
\[ r_i [\vec{x} \mapsto \vec{t^*}, y \mapsto u^*] ~~\reducesto_h^*~~ 
   \lambda \vec{f}.\,\caseof{{T'_j}^*\vec{q'}^*}{\cdots} \]
where $\vec{f} \vdash {T'_j}^* \vec{q'}^*$.
But $r_i[\vec{x} \mapsto \vec{t^*}, y \mapsto u^*] \succeq {z_i^\eta}^* \approx \lambda \vec{w}.0$,
so $\dang{r_i[\vec{x} \mapsto \vec{t^*}, y \mapsto u^*]}$ can only be the procedure $\lambda \vec{f}.0$.
Thus the subterm $\caseof{{T'_j}^*\vec{q'}^*}{\cdots}$ above evaluates to $0$, and so 
${T'_j}^*\vec{q'}^*$ itself must evaluate to some numeral, say $m$.
Finally, specializing $\vec{f}$ to $\vec{\bot}$, we have 
$({T'_j}^*[\vec{f} \mapsto \vec{\bot}])(\vec{q'}^*[\vec{f} \mapsto \vec{\bot}]) \reducesto^* m$,
and so ${T'_j}^*[\vec{f} \mapsto \vec{\bot}] \not\approx \bot$.
Since $\vec{T'}^*$ is what defines the section half of the pseudo-retraction associated with 
$\vec\tau \isplit \vec{\tau'}$, this amounts to saying that this quasi-retraction is non-strict.

We have thus obtained a contradiction in Subcase~2.3,
and the argument for Case~2 is now complete.

\vspace*{1.5ex}
\emph{Case 3}: $\alpha$ is infinite but not of the form $\omega^\delta.k$.
Then writing $\alpha$ in Cantor form as $\alpha_0 + \cdots + \alpha_{r-1}$, we see that the $\alpha_i$
are not all equal; hence writing $\alpha_0$ as $\omega^\delta$ where $\delta>0$,
we may express $\alpha$ as $\omega^\delta.k + \beta$ where $0 < \beta < \omega^\delta$
and $\beta$ has Cantor normal form $\alpha_k + \cdots + \alpha_{r-1}$.
In terms of types, this means we may express $\sigma = \sigma_\alpha$ (up to a trivial isomorphism)
as $(\vec{\tau} \arrow \nat)^k \times \vec{\pi}$, where $R(\vec{\tau})=\delta$ and $R(\vec\pi)=\beta$.
Since $R(\sigma') < \alpha$, we may also express $\sigma'$ as
$(\vec{\tau} \arrow \nat)^{k'} \times \vec{\pi'}$ for some $k' \leq k$, 
where $R(\vec{\pi'}) < \omega^\delta$.
We are thus supposing that we have a quasi-retraction
\[  (\vec\tau\arrow\nat)^k \times \vec{\pi} ~~\isplit~~ 
     ((\vec\tau\arrow\nat)^{k'} \times \vec{\pi'}) \times \vec{\rho}  \]
where $\omega^d \gg R(\vec\rho)$, and we wish to show this is left-strict w.r.t.\
$((\vec\tau\arrow\nat)^{k'} \times \vec{\pi'})$.

We first show that we must have $k'=k$. If not, then coding $\vec{\pi'} \times \vec{\rho}$
as a retract of $(\vec\tau\arrow\nat)$ by Proposition~\ref{ranks-match-prop},
we could obtain a non-strict pseudo-retraction $(\vec\tau\arrow\nat)^k \pseudo (\vec\tau\arrow\nat)^k$.
As in the proof of Proposition~\ref{lemma-implies-theorem}, we could then build a quasi-retraction
$(\vec\tau\arrow\nat)^k \isplit (\vec\tau\arrow\nat)^k$ with this as its associated pseudo-retraction,
contradicting the induction hypothesis for $\omega^\delta < \alpha$.
We thus have $k'=k$; and since $R(\sigma') \leq R(\sigma)$, it also follows that
$R(\vec{\pi'}) \leq R(\vec{\pi})$.

Suppose now that our quasi-retraction is witnessed by procedures
\begin{eqnarray*} 
   z : (\vec\tau\arrow\nat)^k \arrow \vec{\pi} \arrow \nat & \vdash &
   u  \equiv \lambda \vec{f'}\vec{g'}\vec{h}.d ~:~  (\vec\tau\arrow\nat)^k \arrow \vec{\pi'} \arrow \vec{\rho} \arrow \nat \;, \\
   x :  (\vec\tau\arrow\nat)^k \arrow \vec{\pi'} \arrow \vec{\rho} \arrow \nat & \vdash &
   r \equiv \lambda \vec{f}\vec{g}.e  ~:~  (\vec\tau\arrow\nat)^k \arrow \vec{\pi} \arrow \nat
\end{eqnarray*}
(where $\vec{f},\vec{f'}: (\vec{\tau}\arrow\nat)^k$, $\vec{g}:\vec{\pi}$, $\vec{g'}:\vec{\pi'}$ 
and $\vec{h} : \vec{\rho}\,$), such that we have
\[ z ~\vdash~ r [x \mapsto u] ~~\reducesto_h^*~~ 
   \caseof{z P_0 \ldots P_{k-1} \vec{Q}}{\cdots} \]
where $\vec{P}^* \succeq \vec{f}^{\,\eta}$ and $\vec{Q}^* \succeq \vec{g}^{\,\eta}$.
Here $e$ must have the form $\caseof{x\vec{a}\vec{b}\vec{c}}{\cdots}$ for some 
$x,\lambda\vec{f},\lambda\vec{g} \;\vdash\, \vec{a} : (\vec\tau\arrow\nat)^k,\, \vec{b} : \vec{\pi'},\, \vec{c} : \vec{\rho}$.
Furthermore, by Proposition~\ref{simple-qr-prop} we may assume without loss of generality 
(and without effect on the associated pseudo-retraction) that our quasi-retraction is \emph{simple}, 
so that $e$ contains no occurrences of $x$ other than this head one.


Denote the associated pseudo-retraction as
\[ (\ang{\ang{t',s'},\ang{u',v'}},\,r') ~~:~~ 
   (\vec\tau\arrow\nat)^k \times \vec{\pi} ~~\pseudo~~ 
   ((\vec\tau\arrow\nat)^k \times \vec{\pi'}) \times (\vec{\rho} \times \vec{\upsilon}) \;, \]
where $R(\vec{\upsilon}) < R(\vec{\tau})$. 
Here $t'$ arises from the meta-terms $\vec{T'}$ which are substituted for the 
bound variables $\vec{f'}$ of $u$ (of types $(\vec\tau\arrow\nat)^k$)
in the course of the above head reduction: specifically, $t'$ is defined by $\vec{T'}^*$,
where $^* = [z \mapsto \lambda\vec{w}.0]$.
Our task is to show that this pseudo-retraction is left-strict w.r.t.\ $(\vec\tau\arrow\nat)^k \times \vec{\pi}$.

\emph{Claim 4:} For all $\vec{q} \in \SP^0(\vec{\pi})$ we have $t'[\bot^k,\vec{q}\,] \approx \bot^k$.

\emph{Proof:} Take any $\vec{q} \in \SP^0(\vec{\pi})$.
By fixing the value of the $\vec{\pi}$ component at $\vec{q}$ on the left side of the 
above pseudo-retraction, we obtain a pseudo-retraction 
\[ (\ang{t'_q,\ldots},\,r'_q) ~:~ (\vec\tau \arrow \nat)^k \pseudo 
   (\vec\tau \arrow \nat)^k \times (\vec{\pi'} \times \vec\rho \times \vec\upsilon) \;. \]
As noted earlier, it is easy to present any given pseudo-retraction as arising from a quasi-retraction,
so we may obtain a quasi-retraction
$(\vec\tau \arrow \nat)^k \isplit (\vec\tau \arrow \nat)^k \times (\vec{\pi'} \times \vec\rho \times \vec\upsilon)$.
By the induction hypothesis for $\omega^\delta.k$, this must be left-strict w.r.t.\ $(\vec\tau \arrow \nat)^k$.
In other words, the above pseudo-retraction $(\ang{t'_q,\ldots},r'_q)$ is left-strict w.r.t.\ 
$(\vec\tau \arrow \nat)^k$, which establishes Claim~4.

We now show that the terms $z \vdash u$ and $x \vdash r$ can be modified to yield terms
\begin{eqnarray*} 
   z' :  \vec{\pi} \arrow \nat & \vdash & u' \equiv \lambda\vec{g'}\vec{h}.d' ~:~ \vec{\pi'} \arrow \vec{\rho} \arrow \nat \;, \\
   x' :  \vec{\pi'} \arrow \vec{\rho} \arrow \nat & \vdash & r' \equiv \lambda\vec{g}.e' ~:~ \vec{\pi} \arrow \nat
\end{eqnarray*}
that constitute a quasi-retraction $\vec{\pi} \isplit \vec{\pi'} \times \vec{\rho}$, essentially by specializing the
components of type $\vec{\tau}\arrow\nat$ on both sides to $\bot$.
Specifically, we set 
\[ D' ~=~ d\,[\vec{f'} \mapsto \vec{\bot},\, z \mapsto \lambda \vec{f}\vec{g}.\, z' \vec{g}^{\,\eta}] \;,
   ~~~~~~
   E' ~=~  e\,[\vec{f} \mapsto \vec{\bot},\, 
                    x \mapsto \lambda \vec{f'}\vec{g'}\vec{h}.\, x'\vec{g'}^{\,\eta}\vec{h}^{\,\eta}] \;, \]
and then take $d' =\, \dang{\!D'\!}$, $e' =\, \dang{\!E'\!}$.
It will also be useful to set $U' = \lambda\vec{g'}\vec{h}.D'$ and $R' = \lambda\vec{g}.E'$, 
so that $z' \vdash U'$ and $x' \vdash R'$.

\emph{Claim 5:} $z' \vdash u'$ and $x' \vdash r'$ constitute a quasi-retraction
$\vec{\pi} \isplit \vec{\pi'} \times \vec{\rho}$.

\emph{Proof:} This is basically a tedious syntactic verification.
To show that $r'[x' \mapsto \vec u']$ has a head normal form of the appropriate kind,
we will first compare the head reduction sequence for $R'[x' \mapsto U']$ with that for $r[x \mapsto u]$.
Recall that $r$ has the form $\lambda\vec{f}\vec{g}.\,\caseof{x \vec{a} \vec{b} \vec{c}}{\cdots}$,
where by assumption this is the critical occurrence of $x$, and $x$ does not appear in $\vec{a},\vec{b}$.
We therefore have 
\begin{eqnarray*}  
    r[x \mapsto u] & \equiv & 
    \lambda \vec{f}\vec{g}.\; \caseof{u \vec{a} \vec{b} \vec{c}}{\cdots}  \\
    & \reducesto_h &
    \lambda \vec{f}\vec{g}.\; \caseof {d\, [\vec{f'} \mapsto \vec{a},\, \vec{g'} \mapsto \vec{b},\, \vec{h} \mapsto \vec{c}\,]}{\cdots}
\end{eqnarray*}
so that the meta-terms $\vec{T'}$ mentioned earlier are just $\vec{a}$
(and incidentally $z$ does not appear free in $\vec{T'}$).
We claim, furthermore, that none of these occurrences of terms $\vec{a}$ or their residuals ever come into
head operator position in the course of the subsequent head reduction.
This is because 
$\vec{a}[\vec{f} \mapsto \vec{\bot}] = \vec{T'}[\vec{f} \mapsto \vec{\bot}] = \vec{T'}^*[\vec{f} \mapsto \vec{\bot}] \approx \vec{\bot}$,
so each $a_i$ either has no head normal form or has one with some head variable $f_j$;
hence the same would be true for any meta-term headed by an $a_i$, which is not compatible with
$u \vec{a} \vec{b} \vec{c}$ reducing to a term headed by $z$.

Now let us examine the reduction for $R'[x' \mapsto U']$.
Writing $^\star$ for the substitution $[\vec{f} \mapsto \vec{\bot}]$, we have:
\begin{eqnarray*}
   R'[x' \mapsto U'] & \equiv & 
   \lambda \vec{g}.\; \caseof{(\lambda \vec{f'}\vec{g'}\vec{h}.\, U' \vec{g'}^{\,\eta}\vec{h}^{\,\eta})\, \vec{a}^{\,\star} \vec{b}^{\,\star} \vec{c}^{\,\star}}{\cdots} \\
   & \reducesto_h & \lambda \vec{g}.\; \caseof{U' \vec{b}^{\,\star\eta} \vec{c}^{\,\star\eta}}{\cdots}  \\
   & \reducesto_h & \lambda \vec{g}.\; \caseof{D' [\vec{g'} \mapsto \vec{b}^{\,\star\eta},\, \vec{h} \mapsto \vec{c}^{\,\star\eta}]}{\cdots}  \\
   & \equiv & \lambda \vec{g}.\; \caseof{d^\star [\vec{g'} \mapsto \vec{b}^{\,\star\eta}, 
          \vec{h} \mapsto \vec{c}^{\,\star\eta}, z \mapsto \vec{f}\vec{g}. z'\vec{g}^{\,\eta}]}{\cdots}
\end{eqnarray*}

Comparing the two reductions, we see that
$d^\star [\vec{g'} \mapsto \vec{b}^{\,\star\eta}, \vec{h} \mapsto \vec{c}^{\,\star\eta}, z \mapsto \vec{f}\vec{g}. z'\vec{g}^{\,\eta}]$
may be obtained from $d [\vec{f'} \mapsto \vec{a},\, \vec{g'} \mapsto \vec{b},\, \vec{h} \mapsto \vec{c}\,]$
by first replacing certain occurrences of subterms $\vec{a}$ by $\bot$, then applying the specialization
$^\dag = [\vec{f} \mapsto \bot,\, z \mapsto \vec{f}\vec{g}.z'\vec{g}^{\,\eta}]$.
Since, as we have observed, none of the $\vec{a}$ occurrences in question ever come into head position,
it is clear that the subsequent head reduction of $r[x \mapsto u]$ to $\caseof{z \vec{P} \vec{Q}}{\cdots}$
will be matched by a corresponding head reduction of $R'[x' \mapsto U']$ to some meta-term
$\caseof{(\vec{f}\vec{g}.z'\vec{g}^{\,\eta}) \vec{P}^{\,\bullet} \vec{Q}^{\,\bullet}}{\cdots}$, which
in turn head-reduces to $\caseof{z' \vec{Q}^{\,\bullet\eta}}{\cdots}$.
Here $\vec{Q}^{\,\bullet}$ will be some sequence of terms obtained from $\vec{Q}$ by replacing 
certain subterms $\vec{a}$ by $\bot$ then applying $^\dag$.
But since $\vec{a}[\vec{f} \mapsto \vec{\bot},\, z \mapsto \lambda \vec{w}.0] \equiv \bot$,
and the substitution $z \mapsto \vec{f}\vec{g}.z'\vec{g}^{\,\eta}$ 
followed by $z' \mapsto \lambda \vec{w}.0$ is equivalent to $z \mapsto \lambda \vec{w}.0$, we have
\[ \vec{Q}^{\,\bullet\eta}[z' \mapsto \lambda\vec{w}.0] ~\approx~ 
    \vec{Q}^{\,\bullet}[z' \mapsto \lambda\vec{w}.0] ~\approx~
    \vec{Q}[\vec{f} \mapsto \vec{\bot},\,\vec{z \mapsto \lambda\vec{w}.0}] ~\preceq~ 
    \vec{g}^{\,\eta}[\vec{f} \mapsto \vec{\bot}] ~\equiv \vec{g}^{\,\eta} \]
We have thus shown that $R'[x' \mapsto U'] \reducesto_h^* \caseof{z' \vec{Q}^{\,\bullet\eta}}{\cdots}$
where $\vec{Q}^{\,\bullet\eta} \succeq \vec{g}^{\,\eta}$, and it follows that also
\( r'[x' \mapsto u'] ~~\reducesto_h^*~~ \caseof{z' \vec{Q}^{\,\circ}}{\cdots} \)
for some $\vec{Q}^{\,\circ}$ with $\vec{Q}^{\,\circ} \succeq \vec{g}^{\,\eta}$.
This exhibits a quasi-retraction $\vec{\pi} \isplit \vec{\pi'} \times \vec{\rho}$,
establishing Claim~5.

Recalling now that $R(\vec{\pi}) < \alpha$ and that $R(\vec{\pi'}) \leq R(\vec{\pi})$ 
and $R(\vec{\pi}) \gg R(\vec{\rho})$,
the induction hypothesis tells us that this quasi-retraction must be left-strict w.r.t.\ $\vec{\pi'}$.
This amounts to saying that $b^{\,\star}[\vec{g} \mapsto \vec{\bot}] \approx \vec{\bot}$,
i.e.\ that $b[\vec{f},\vec{g} \mapsto \vec{\bot}] \approx \vec{\bot}$.
But we have already seen that $a[\vec{f} \mapsto \vec{\bot}] \approx \vec{\bot}$,
and this means that the pseudo-retraction $(\ang{\ang{t',s'},\ang{u',v'}},r')$ is left-strict
w.r.t.\ $(\vec{\tau}\arrow\nat)^k \times \vec{\pi'}$,
since $t',s'$ are defined by $\vec{a},\vec{b}$ respectively so that
$\ang{t',s'}[\vec{\bot},\vec{\bot}] = (\vec{\bot},\vec{\bot})$.
This concludes the argument for Case~3.

\vspace*{1.5ex}
The proof of Lemma~\ref{main-lemma} is now complete,
and Theorem~\ref{main-theorem} follows by Proposition~\ref{lemma-implies-theorem}.

\vspace*{2.5ex}
Evidently, the above analysis depends on working with just a single base type $\nat$.
The picture will of course become more complex if base types such as $\unit$ and $\bool$ 
(unit and booleans) are added.
Here the encodability relation will of course no longer be total, since for example 
neither $\nat$ nor $\nat\arrow\bool$ can be encoded in the other.
Nevertheless, we expect that our hierarchy for types over $\nat$ to play a central
role in mapping out the encodability relation in more general settings.

\end{document}